\pdfoutput=1
\documentclass[a4paper,11pt]{article}
\usepackage{pos}

\usepackage{amsmath}
\usepackage{amssymb}

\usepackage{newtxmath}
\usepackage{subcaption}
\usepackage{float}
\usepackage{braket}
\usepackage{wrapfig}
\usepackage{bbold}
\usepackage{hyperref}

\usepackage{graphicx} 
\usepackage{booktabs} 
\usepackage{subcaption}

\title{$K \pi$ scattering at physical pion mass using distillation}

\author*[a]{Nelson Pitanga Lachini}
\author[a,b]{Peter Boyle}
\author[a]{Felix Erben}
\author[a]{Michael Marshall}
\author[a]{Antonin Portelli}

\affiliation[a]{School of Physics and Astronomy, University of Edinburgh, Edinburgh EH9 3JZ, UK,}
\affiliation[b]{Brookhaven National Laboratory, Upton, NY 11973, USA}

\emailAdd{nelson.lachini@ed.ac.uk}

\abstract{Scattering at physical pion mass is still an exploratory field in lattice QCD. This generally involves the extraction of excited states through multi-particle correlators on systems with resonances. In that context, distillation has been demonstrated to be effective both as a smearing kernel and a computational tool. Motivated by the study of the smearing profile of the distillation operator, we compare stochastic and exact distillation cases for different numbers of Laplacian eigenvectors using a RBC-UKQCD $N_f=2+1$ domain-wall fermion lattice with a physical pion mass.}

\FullConference{%
 The 38th International Symposium on Lattice Field Theory, LATTICE2021
  26th-30th July, 2021
  Zoom/Gather@Massachusetts Institute of Technology
}

\begin{document}
\maketitle

\section{Introduction}

Lattice QCD is formulated on a Euclidean space and simulations always take place on finite volumes, which means that there is no way of obtaining infinite-volume scattering information directly from the lattice. We only have access to the Euclidean spectrum of the theory. Nevertheless, a connection between the spectrum of a finite-volume Euclidean theory and the scattering amplitude of the correspondent infinite-volume Minkowski theory was established already in the 1980s \cite{Luscher1986,Luscher1986a} and was further developed over the years \cite{Rummukainen1995,Kim2005a}. Moreover, lattice QCD is able to address hadronic resonances by studying the scattering phase shift obtained from such finite-volume analysis \cite{Luscher1991a}.

The number of lattice studies of hadronic resonances has increased over the last years but still is in an early stage \cite{Briceno2018}. As we approach physical pion masses at fixed $m_\pi L \sim 4$, the number of energy levels in the regime of elastic scattering decreases. This further motivates the inclusion of moving frames on the calculation in order to satisfactorily constrain phase shift models. In particular, we will be interested in studying $P$-wave $K \pi$ scattering and the $K^*(892)$ resonance at a physical pion mass. This calculation was addressed previously at higher pion masses by other collaborations \cite{Rendon2020,Brett2018,Prelovsek2013a,Bali2016}.

Another clear difficulty of scattering studies at physical pion mass is its computational cost. As we decrease $m_{\pi}$, the physical spatial extension $L$ must be increased in order to keep exponentially suppressed finite-volume effects under control. For scattering studies, it is crucial to have access to a reasonable operator basis in order to use variational analysis methods, and for this we use distillation to smear quark fields and compute lattice correlators \cite{Peardon2009,Morningstar2011}. 

After looking at the smearing radius of the distillation operator, we have done a tuning study across different setups using $9$ configurations of the RBC-UKQCD $N_f=2+1$ domain-wall fermion lattice with $m_{\pi}\approx 139$ MeV and $m_K\approx 499$ MeV and a $48^3 \times 96$ volume \cite{Blum2016}. Inversions of the Dirac operator were done on every $8$th time slice, and extended in exact distillation to every $2$nd time slice. We observe that for our case, opting for exact (non-stochastic) distillation is an optimal choice and we outline the variational analysis setup in its early stage. We showcase a preliminary variational analysis on the exact distillation data by solving a simple $5 \times 5$ GEVP for a $K\pi$ system in the rest frame.

We use Grid as the data parallel C++ library for the lattice computations and Hadrons as the workflow management system \cite{Boyle2015,Hadrons2020}. The computation of meson fields efficiently and at manageable storage cost demanded writing of dedicated code within Grid and Hadrons, which are being documented and released as open-source and free software.  The previous version of the distillation code within Grid and Hadrons was first showcased in Ref. \cite{Boyle2019}.

\section{Distillation}

In order to compute the correlator data necessary to extract the finite-volume energy spectrum, we use the so-called distillation method \cite{Peardon2009}. Distillation involves a combination of link smearing and $3D$-Laplacian (Lap) quark smearing, which are both gauge covariant by construction. Suppose a lattice of spatial volume $N^3$, time extension $N_t$, number of Dirac components $N_D$ and number of colors $N_c$. Given the gauge-covariant $3D$-Laplacian \cite{Peardon2009} eigenvalues and eigenvectors on a certain time slice, namely
\begin{equation}
- \nabla^2 v_k(t) = \lambda_k (t) v_k(t), \quad k=1, 2, \ldots, \qquad 0 < \lambda_1 < \lambda_2 < \ldots \ .
\end{equation}
we define the smearing operator for distillation as a projector onto the low-mode subspace of $- \nabla^2$, i.e.
\begin{equation}
	\label{distillationkernel}
  \mathcal{S} (t)  = \sum_{k=1}^{N_{\mathrm{vec}}} v_k (t)  v_k (t)^{\dagger}.
\end{equation}  
The distillation operator defines the smeared quark field through $\psi(t) \to \mathcal{S} (t) \psi(t)$. The suppression of short-distance modes is desirable as they are deemed to contribute less to low-energy physical signals in hadron correlation functions.

Given the Dirac operator $D$, smearing the quark field effectively defines the distilled quark propagator $S$ as
\begin{align}
\label{distilledprop}
& S_{a b, \alpha \beta}(x,y) \equiv \left[\mathcal{S} D^{-1} \mathcal{S}^{\dagger}\right]_{a b , \alpha \beta}(x,y) = \sum_{\tilde{\bold x},\tilde a} \mathcal{S}_{a\tilde a}(\bold x , \tilde{\bold x} ; t_f) \sum_{\tilde{\bold y},\tilde b} D_{\alpha \beta,\tilde a \tilde b}^{-1}(\tilde{\bold x},t_f; \tilde{\bold y} , t) (\mathcal{S}^{\dagger})_{\tilde b b}(\tilde{\bold y} , \bold y ; t) \\ \nonumber
& = \sum_k \sum_l v_{ak}(\bold x ;t_f) \left[\sum_{\tilde{\bold x},\tilde a} v_{\tilde a k} (\tilde{\bold x} ;t_f)^* \sum_{\tilde{\bold y},\tilde b} D_{\alpha \beta,\tilde a \tilde b}^{-1}(\tilde{\bold x},t_f; \tilde{\bold y} , t) v_{\tilde b l}(\tilde{\bold y} ; t)\right] v_{b l}(\bold y; t)^*,
\end{align}
where the term between square brackets is the perambulator $\tau_{\alpha \beta,kl}(t_f , t)$. This is what is referred to as \textit{exact} distillation.

The perambulator gives us access to all spatial entries of the propagator between $t_f$ and $t_0$. Due to the projection of the space-color subspace into the distillation subspace, the number of inversions needed to access a perambulator is way smaller than the number of inversions for an explicit full propagator (all color-positions to all color-positions), namely $4 N_{\mathrm{vec}} \sim \mathcal{O}(10^2)$ against $12 N^3 \sim \mathcal{O}(10^6)$ per source and sink time slice, quark flavor and configuration (typically for $\mathcal{O}(10)$ time slices and $\mathcal{O}(10^2)$ configurations).

\subsection{Stochastic and Diluted Distillation}
Instead of evaluating $S$ exactly using distillation, we can stochastically evaluate the perambulators in the spin-time-Lap subspace, which can be efficient when the noise introduced is not greater than the gauge noise \cite{Morningstar2011}. For that, we define the spin-time-Lap noises $\eta = [\eta^r_{\alpha k}(t_f)]$, which obey
\begin{equation}
\Braket{\eta^{r}_{\alpha k}(t_f)}_{\eta} = 0, \qquad
\label{noiseorthonormality}
\Braket{\eta_{\alpha k}^{r}(t_f) \eta_{\beta l}^{s}(t)}_{\eta} = \delta_{rs} \delta_{kl} \delta_{\alpha \beta} \delta(t_f-t),
\end{equation}
where $r,s$ are noise hit indices and the expectation value is given by the usual definition $\Braket{O}_{\eta} = \lim_{N_{\eta}\to\infty} \frac{1}{N_{\eta}} \sum_{r=1}^{N_{\eta}}O^r $. 

It is also desirable to reduce the variance of such an estimation by introducing exact zeros through dilution. We partition a single dilution space $I$ as in
\begin{equation}
\label{dilutionspacepartitions}
I = \bigcup_{i=1}^{N_i} I^i, \qquad i_1 \neq i_2 \Rightarrow I^{i_1} \cap I^{i_2} = \varnothing, \quad I^i \neq \varnothing,
\end{equation}
where $I^i$ is a dilution partition labeled by $i$, which is a subset of $I$. There are no overlaps between different partitions and no partition is empty. Furthermore, when combining different dilution spaces, we demand the overlap between them to also be zero. The standard way of projecting different objects onto the various partitions is to define dilution projectors in spin-time-Lap spaces, respectively, $P^{S} , P^{T}$ and $P^{L}$, and the collective projector $P^{d} \equiv P^{S} P^{T} P^{L}$, where $d$ is a compound index containing $S, T, L$. The projectors have the usual properties 
\begin{equation}
(P^{d})^2 = P^{d}, \qquad \sum_{d} P^{d} = \mathbb{1},
\end{equation} 
where $\mathbb{1}$ here is the identity operator in the whole dilution space considered.

Using stochastic noises and dilution projectors, we can factorise the propagator into two vectors, $\varphi,\varrho$, as in
\begin{equation}
S_{a b}^{\alpha \beta}(\bold x,t_f;\bold y,t) = \Bigg \langle \sum_{d} \varphi^{d}_{a \alpha}(\bold x , t_f) \varrho^{d}_{b \beta}(\bold y,t)^* \Bigg \rangle_{\eta} = \Big \langle S_{a b, \alpha \beta}(x,y) \Big \rangle_{\eta},
\end{equation}
where the source vector is
\begin{equation}
	\varrho^{d}_{b \beta}(\bold y,t) \equiv \sum_l v_{b l}(\bold y; t) \sum_{\sigma,n,w_0} P^{d}_{\beta \sigma,ln}(t,w_0)  \eta_{\sigma n}(w_0) = \sum_l v_{b l}(\bold y; t) \left(P^{d} \eta \right)_{\beta l}(t)
\end{equation}
and the correspondent solution vector is
\begin{equation}
	\varphi^{d}_{a \alpha}(\bold x , t_f) \equiv \sum_{k} v_{ak}(\bold x ;t_f) \Bigg[ \sum_{\tilde \beta,\tilde{t}} \sum_{\tilde{\bold x},\tilde a} v_{\tilde a k} (\tilde{\bold x} ;t_f)^* \sum_{\tilde{\bold y},\tilde b} D_{\alpha \tilde{\beta},\tilde a \tilde b}^{-1}(\tilde{\bold x},t_f; \tilde{\bold y} , \tilde{t}) \varrho^{d}_{\tilde b \tilde{\beta}}(\tilde{\bold y},\tilde{t}) \Bigg].
\end{equation} 

In matrix notation, we can express a simple $2$-point correlation function as
\begin{align}
\label{simplecorrelator}
C(x,y) & = - \Braket{\text{tr}\left[\Gamma S(x,y) \Gamma S'(y,x)\right]}_{\eta} = - \Braket{\sum_{d_1,d_2} \left[\varrho^{d_1}(y)^{\dagger} \Gamma_1 \varphi^{d_2}(y) \ \varrho^{d_2}(x)^{\dagger} \Gamma_2 \varphi^{'d_1}(x)\right]}_{\eta} \nonumber \\
& =  - \Braket{\sum_{d_1,d_2} \left[M_{\Gamma_1}^{d_1,d_2}(\varrho,\varphi;y) \ M_{\Gamma_2}^{d_2,d_1}(\varrho,\varphi';x)\right]}_{\eta}.
\end{align}
The meson fields $M_{\Gamma}^{d_1,d_2}(\varrho,\varphi;x) \equiv \varrho^{d_1}(x)^{\dagger} \Gamma \varphi^{d_2}(x)$ can be further momentum projected to enable computation of momentum-space correlators. Note that we can use the stochastic-diluted notation for exact distillation if we take full dilution in all subspaces, $N_{\eta}=1$ and noise vectors with all entries equal $1$.

\subsection{Smearing Profile and Radius}

Choosing $N_{\mathrm{vec}}$ is not an easy task. In order to keep the $3D$-Lap cut off constant between lattices of different volumes, the number of eigenvectors has to be proportional to the volume. If we were to scale $N_{\mathrm{vec}}$ from an earlier study on a $24^3 \times 64$ RBC-UKQCD lattice \cite{Boyle2019}, we would need $N_{\mathrm{vec}} \sim 500$ low-lying $3D$-Lap eigenvectors, rendering exact distillation prohibitively expensive. On the other hand, having a small enough $N_{\mathrm{vec}}$ can potentially over-smear the quark field, prevent a satisfactory momentum projection and remove excited states that we actually want to study. As this project is tackling a physical-pion mass lattice directly, we decided to tune $N_{\mathrm{vec}}$ independently and report on our findings in this section.  

\begin{figure}[H]
	\centering
    \includegraphics[width=.65\linewidth]{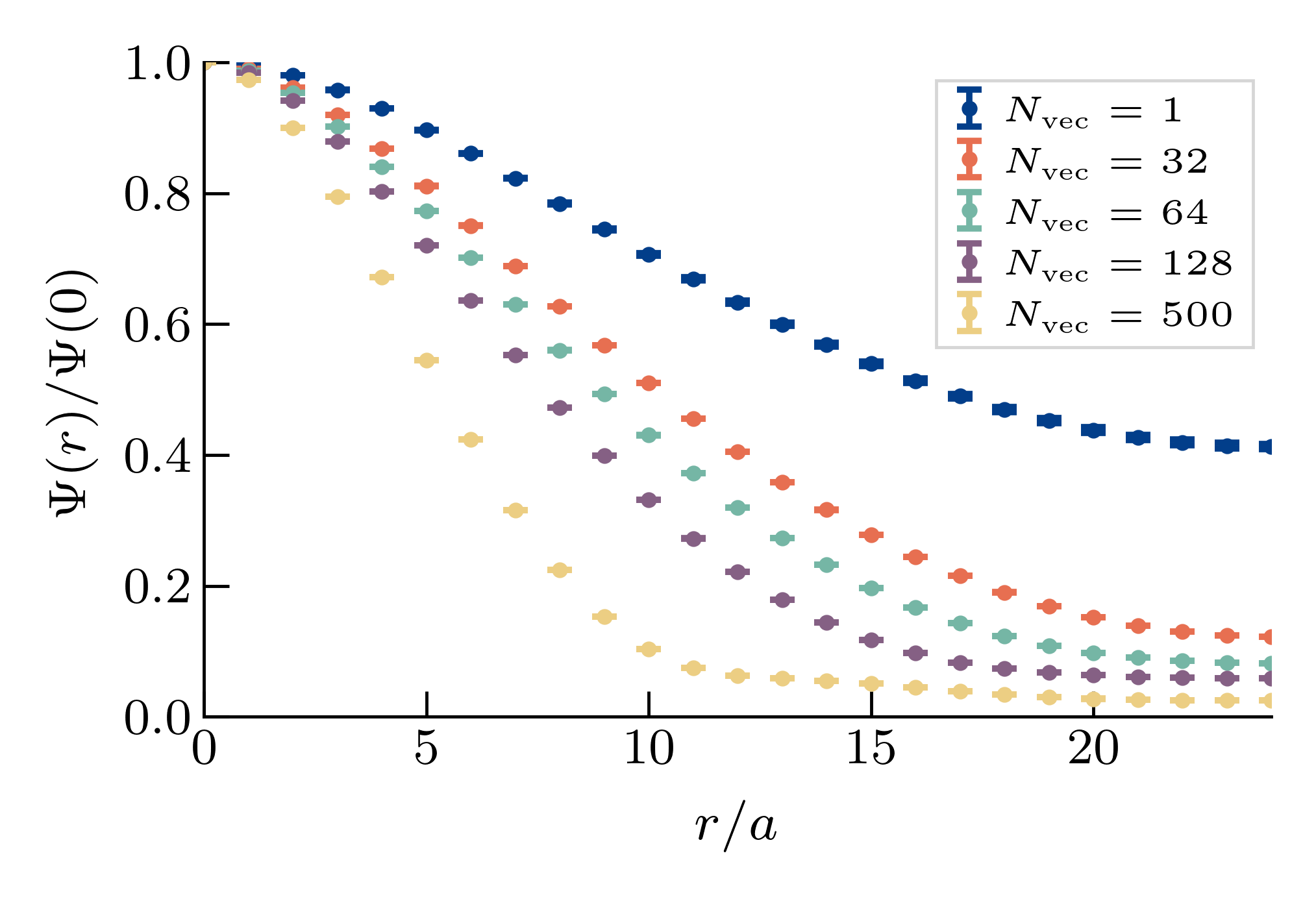}
    \caption{Smearing profile for distillation operator for various $N_{\mathrm{vec}}$. Stout-smearing parameters: $~{\rho=0.2, n=3}$.}
    \label{fig:smearingprofile}
\end{figure}

Given the distillation kernel of Eq. \eqref{distillationkernel} and $r \equiv | \mathbf r |$, we can build a spatial distribution function \cite{Peardon2009}
\begin{equation}
\label{spatialprofile}
\Psi(r) = \sum_{\bold x , t} \sqrt{\text{tr} \ \mathcal{S}_{\bold x , \bold x + \bold r}(t) \mathcal{S}_{\bold x + \bold r , \bold x}(t) }
\end{equation}
which measures the amount of smearing done by the $\mathcal{S}$ operator on the spatial lattices, or in other words, how far is the smeared source from a point source. In Fig. \ref{fig:smearingprofile}, we see that, as $N_{\mathrm{vec}}$ increases, the smearing profile tends to a narrow spatial distribution, which should approximate a delta function for large enough $N_{\mathrm{vec}}$.

\begin{figure}[H]
	\centering
    \includegraphics[width=.70\linewidth]{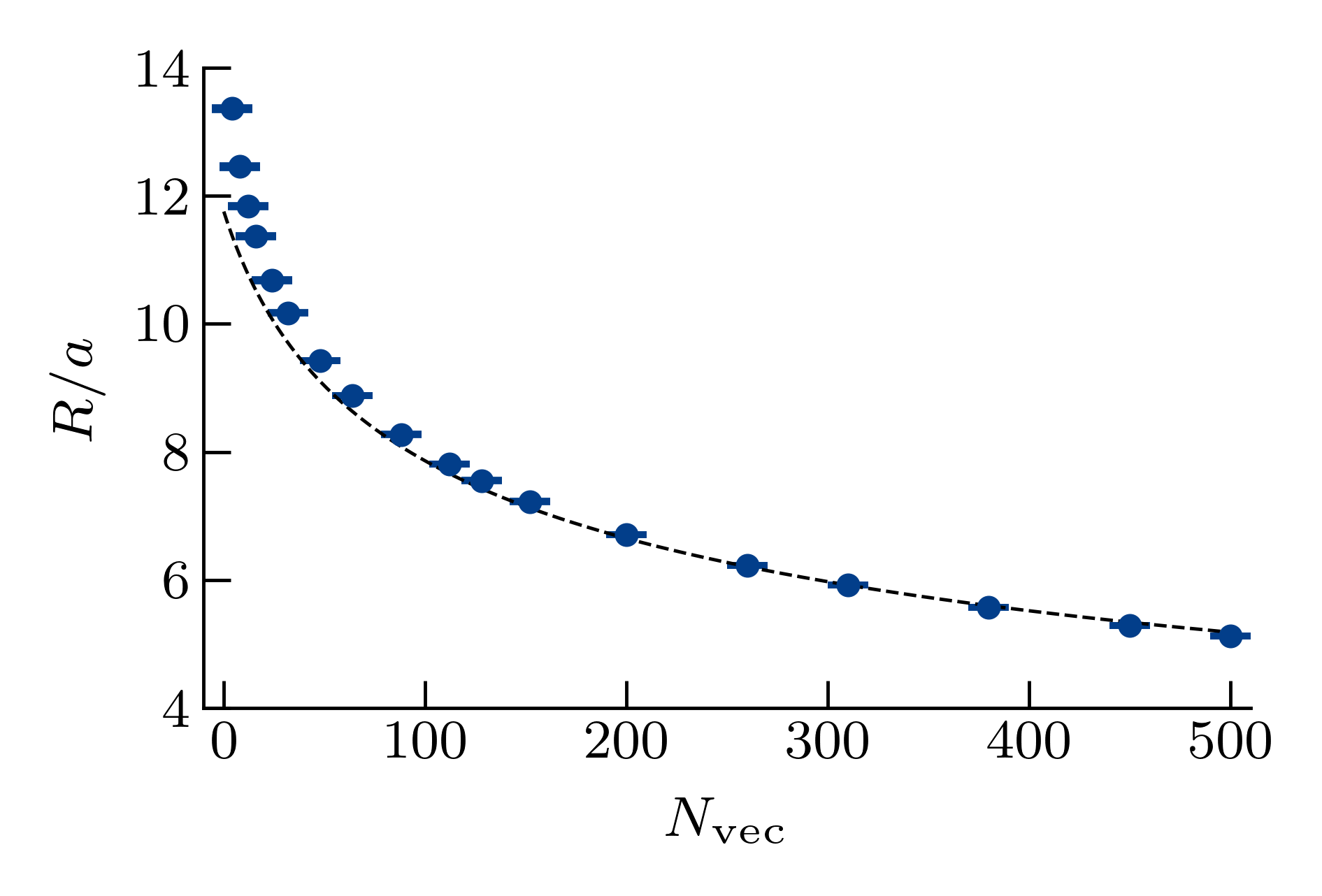}
    \caption{Smearing radius defined in Eq. \eqref{smearingradius} as a function of $N_{\mathrm{vec}}$.  The dashed line is a naive fit to $A( N_{\mathrm{vec}} + B )^C$, yielding $C\approx-0.3$.}
    \label{fig:smearingradius}
\end{figure}

By defining the smearing radius as
\begin{equation}
  	\label{smearingradius}
	\int_0^{R} \Psi(r) d r = 0.682 \int_0^{a L/2} \Psi(r) d r, 
\end{equation}
we can study how the smearing effect of distillation varies with $N_{\mathrm{vec}}$ on this ensemble. In Fig. \ref{fig:smearingradius}, a significant curve flattening starts at $N_{\mathrm{vec}} \sim 60$, showing that the decreasing of the smearing radius in physical units is progressively less effective as $N_{\mathrm{vec}}$ is increased above that regime. Given the cost of a high $N_{\mathrm{vec}}$, this motivated the study of simple observables at the region of $N_{\mathrm{vec}} \sim 100$ using various distillation schemes in Table \ref{distilruns}.

\begin{table}[H]
    \centering
    \begin{tabular}{c|c|c|c|c|c|c|c|}
        \cline{2-8}
        & $LI=4$ & $LI=8$ & $LI=16$     & $LI=32$ & exact & exact  & exact  \\ \hline
        \multicolumn{1}{|c|}{$N_{\mathrm{vec}}$} & $64$   & $64$   & $64,96,128$ & $64$    & $20$  & $40$   & $64$   \\ \hline
        \multicolumn{1}{|c|}{$N_{\mathrm{inv}}$} & $32$  & $64$  & $128$      & $256$  & $80$ & $160$ & $256$ \\ \hline
    \end{tabular}
    \caption{Distillation schemes explored. Full spin and time dilution were used on the stochastic runs and $LI$ corresponds to the number of Laplacian sources on an interlaced scheme \cite{Morningstar2011}. For the exact runs, $N_{\mathrm{inv}} = 4 N_{\mathrm{vec}}$, while for the stochastic ones $N_{\mathrm{inv}} = 2 N_{\eta} LI$ with $N_{\eta}=2$ noises (number of light and strange inversions per configuration per time source).}
    \label{distilruns}
\end{table}

\section{$N_{\mathrm{vec}}$ and Cost Comparison}

A first study on the smearing effect of distillation on simple observables was done using light-strange vector-to-vector correlators, i.e. Eq. \eqref{simplecorrelator} with $\Gamma=\gamma^i$ and one light and one strange propagator, which we call $K^*$-like or vector correlator. 

\subsection{$N_{\mathrm{vec}}$ dependence}
\begin{figure}[H]
	\centering
	\includegraphics[width=.85\linewidth]{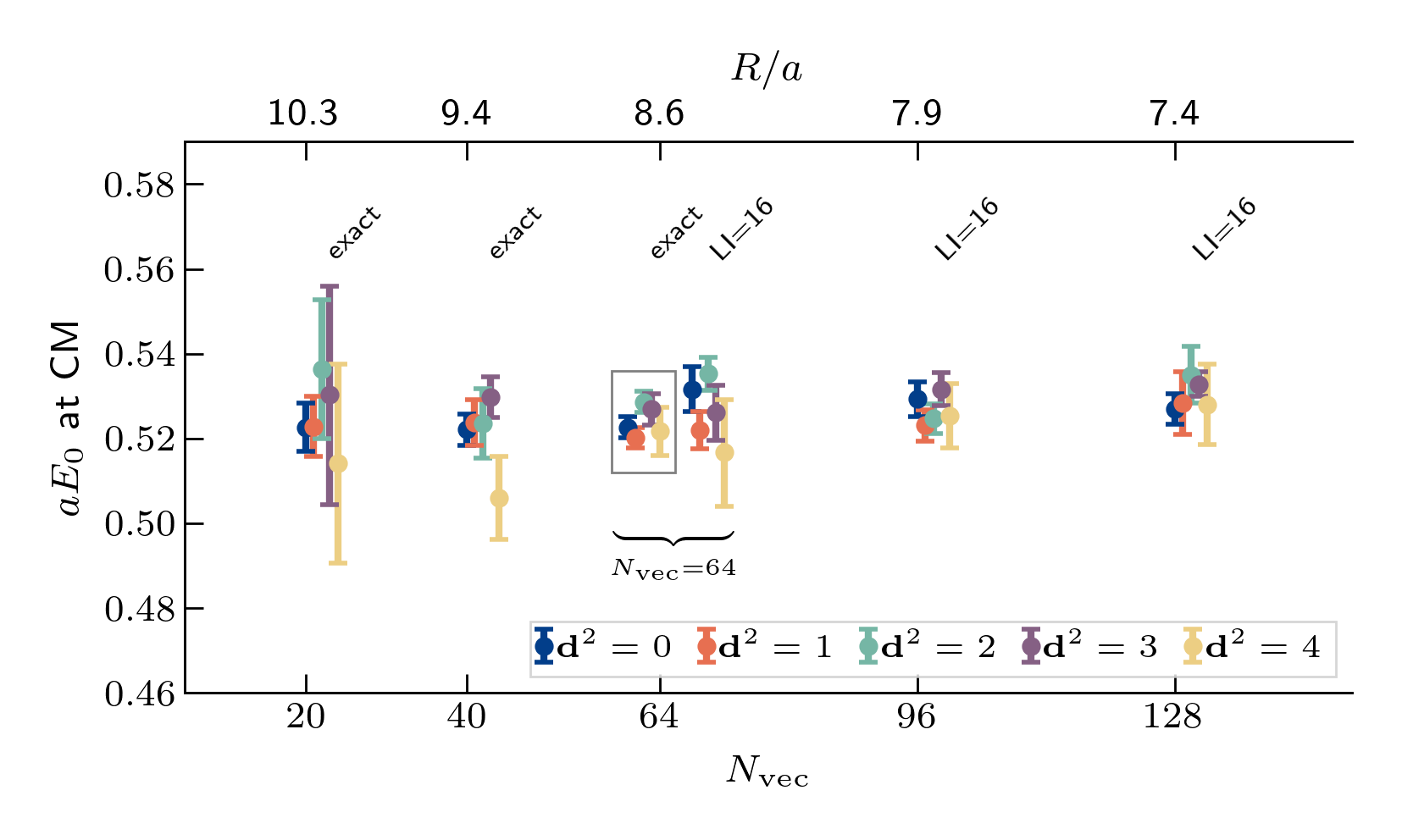}
	\caption{Fitted values of the ground state energy obtained from cosh fits against $N_{\mathrm{vec}}$ (or the smearing radius on top).  Different colors represent different moving frames with total momentum $\mathbf P = 2\pi \mathbf d /L $. Data computed in moving frames ($A_1$ irrep) was boosted to the CM.}
	\label{fig:e0xsmearing}
\end{figure}

Firstly, we study the behavior of the $K^*$-like correlator by computing it at several values of $N_{\mathrm{vec}}=20,40,64,96,128$ without any computational-cost normalization. In Fig. \ref{fig:e0xsmearing}, using a cosh fit function $C(t) = Z_0 \left( e^{- E_0 t} + e^{- (N_t -t)E_0} \right)$, we tested the effectiveness of the momentum projection by comparing $E_0$ at different moving frames through the lattice dispersion relation. The fit ranges were chosen on the exact distillation data with $N_{\mathrm{vec}}=64$.

We observe that for $N_{\mathrm{vec}}\le 64$, there is a spread of different moving frame energies boosted to the CM, which indicates a poor resolution of the momentum-space $K^*$-like operator. Note that the comparison is done between moving frames at each value of $N_{\mathrm{vec}}$, as the cost varies along that axis. Also, by the same argument, we do not see a clear benefit when going above $N_{\mathrm{vec}}=64$, into the stochastic data. 

\begin{figure}[H]
	\centering
	\includegraphics[width=.75\linewidth]{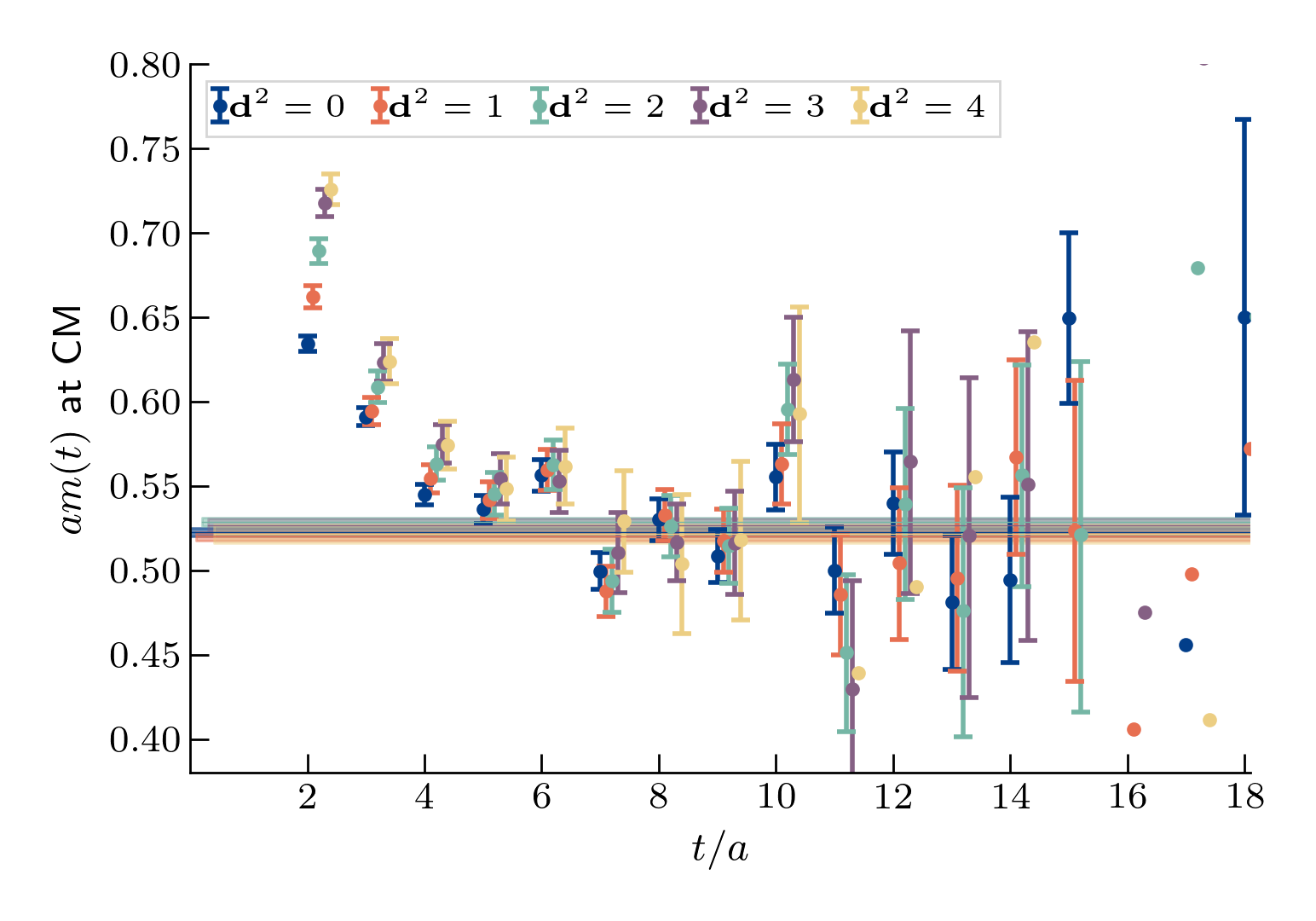}
	\caption{Effective mass at different moving frames boosted to the CM ($N_{\mathrm{vec}}=64$ and exact distillation). The bands represent the $E_0$ fit results contained inside the box in Fig. \ref{fig:e0xsmearing}.}
	\label{fig:zoomeffmassexact}
\end{figure}

The boxed data points refer to exact distillation with $64$ Laplacian eigenvectors, where just gauge noise is present. In Fig. \ref{fig:zoomeffmassexact}, we show the cosh effective mass for that exact setup. We emphasize that the $12$ time translations on each configuration were treated as independent samples, which at low statistics was our best way to estimate errors. Progressive binning of those was performed to check that conclusions did not change dramatically due to correlations. Also, we are only looking at the ground state of a vector correlator, which limits our conclusions about the signal-to-noise of excited states present in this channel.

\subsection{Stochastic and Exact}

Having some level of confidence around the value $N_{\mathrm{vec}}=64$, we can also make a cost-comparison study of the exact and stochastic setups, varying the number of Laplacian-interlaced dilution sources $LI$ and keeping $N_{\mathrm{vec}}$ fixed.

\begin{figure}[H]
	\centering
	\begin{subfigure}{.47\textwidth}
		\centering
		\includegraphics[width=.99\linewidth]{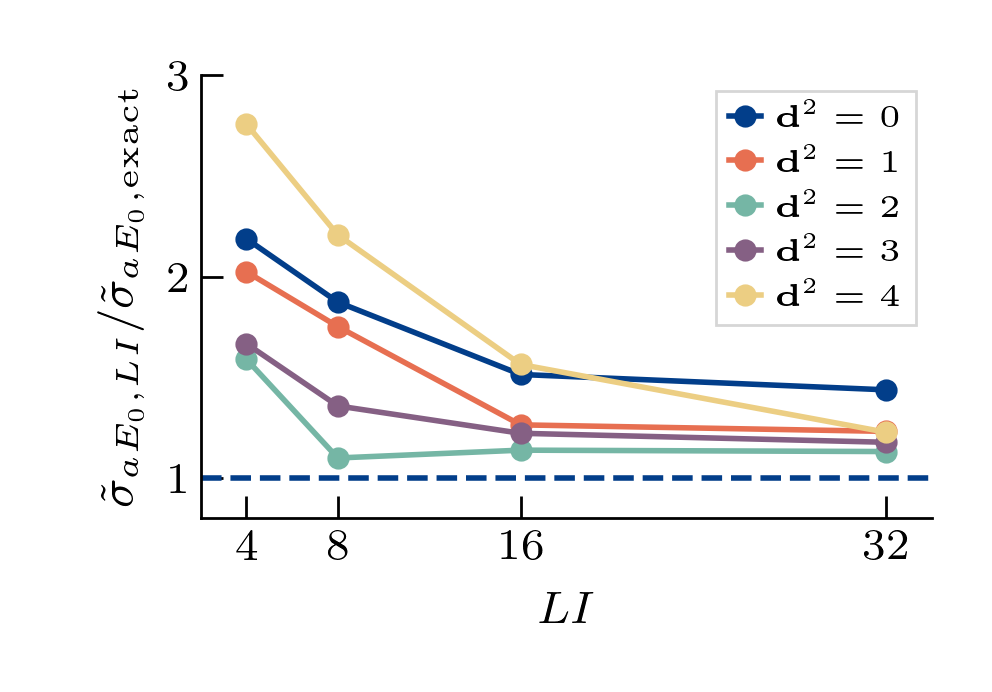}
		\caption{}
		\label{fig:sigmae0xli}
	\end{subfigure}
	\begin{subfigure}{.47\textwidth}
		\centering
		\includegraphics[width=.99\linewidth]{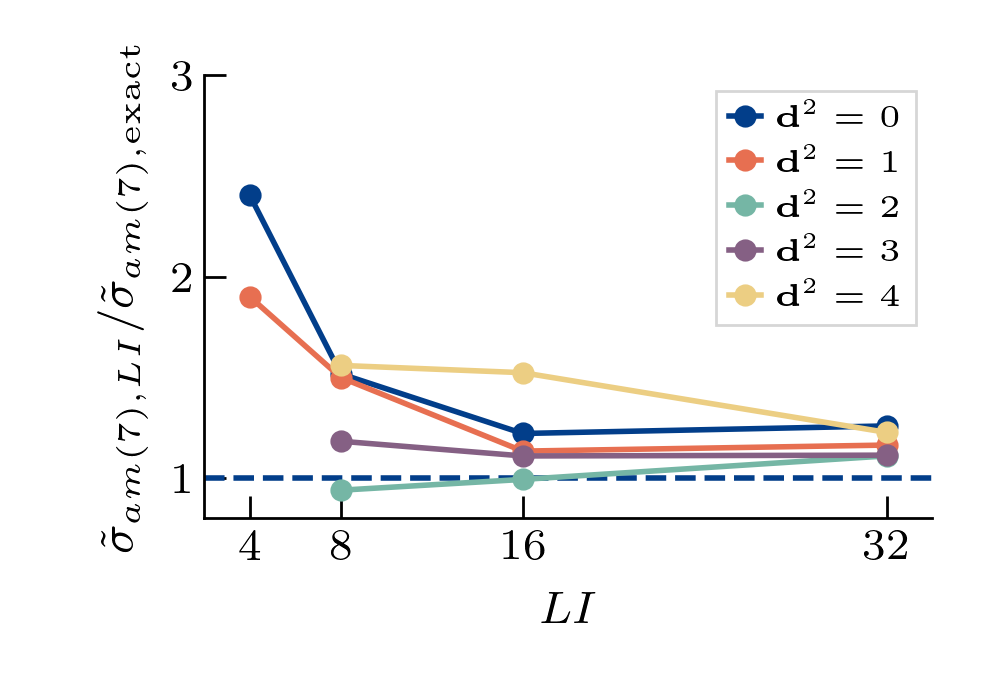}
		\caption{}
		\label{fig:meffxli}
	\end{subfigure}
	\caption{Cost-normalized error ($\tilde \sigma = \sigma \sqrt{N_{\mathrm{inv}}}$) ratio between stochastic and exact distillation for $N_{\mathrm{vec}}=64$ (fitted energy on the left and effective mass at source-sink separation $t=7$ on the right). The dashed line represents the gauge-noise limit (exact distillation).}
	\label{fig:costcomparisonstochexact}
\end{figure}

For that, we can compare the cost-normalized error $\tilde \sigma = \sigma \sqrt{N_{\mathrm{inv}}}$ between results of stochastic distillation and exact, where there is only gauge noise. The dashed lines on Fig. \eqref{fig:costcomparisonstochexact} are a reference for when only gauge noise would be present. As it is expected, lower values of $LI$ tend to be quite above the gauge noise reference. However, as we increase $LI$, the normalized gauge noise limit is not reached even for $LI=32$, where the number of inversions is the same as in exact distillation. On a $K \pi$ scattering workflow, the number of noises would be increased to at least $4$ and the stochastic runs would cost twice as much.

\section{Outlook of Variational Analysis}

To perform a lattice scattering calculation based on the finite-volume formalism, we need to obtain towers of low-lying energies on several moving frames. Defining a variational problem and solving a generalized eigenvalue problem (GEVP) applied to a correlator matrix is a possible way to proceed \cite{Briceno2018}. Multi-hadron correlators are suitable for computing within distillation, as we can combine the same perambulators with different operators and generate correlators with potentially different overlaps to various states.

The previous sections motivated the use of exact distillation and $N_{\mathrm{vec}}=64$. We extended the dataset to have Dirac-operator inversions on every other time slice (for a total of $48$ per configuration on our lattice). This means that the correlators with backtracking lines are defined only on even time slices.

A simple estimation with non-interacting particles shows that even using moving frames up to $|\mathbf P|^2 = 4 (2\pi/L)^2$ only yields $\sim 10$ levels below the inelastic threshold. This means that for each moving frame, a maximum of $2$ energies might be in the elastic regime where the $2$-particle quantisation of the finite-volume formalism is valid. We produced meson fields that are able to construct multi-hadron correlators with total momenta squared up to $4 (2\pi/L)^2$, with the possibility of building up to $ n \times n, n \lesssim 10$ GEVPs depending on the irrep and operator basis.

\begin{figure}[H]
	\centering
	\includegraphics[width=.85\linewidth]{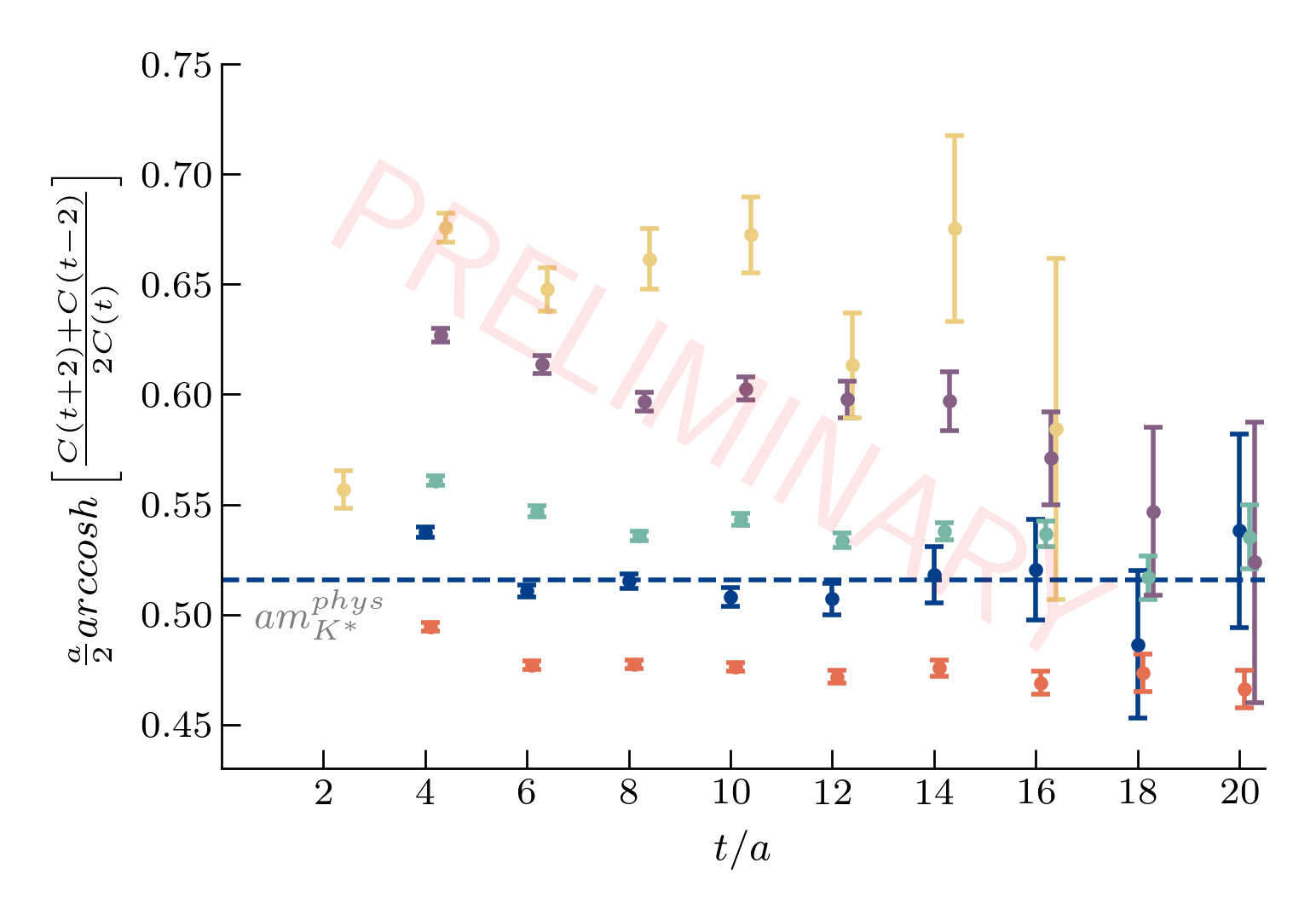}
	\caption{Preliminary rest-frame GEVP on a $5 \times 5$ correlator matrix (fixed-$t_0=2$). The $K \pi$-like operators have back-to-back momenta $\mathbf q = (0,0,1), (0,1,1), (1,1,1), (0,0,2)$. Every $4$ consecutive time translations were binned together to diminish correlations to be treated later. The physical $K^*$ mass ($m_{K^*}^{phys}$) was taken from Ref. \cite{Zyla2020}.}
	\label{fig:gevp}
\end{figure}

We performed a preliminary rest-frame variational analysis ($T_{1u}$ irrep) on a $5 \times 5$ correlator matrix $C(t)$ defined by single-hadron $\left( \bar s \gamma^z l \right)(\mathbf p = 0)$ and multi-hadron $\left( K( \mathbf q) \pi(-\mathbf q) \right)^{I=1/2}$ operators. The GEVP was solved using a simple fixed-$t_0$ method defined by 
\begin{equation}
    C(t) u(t,t_0) = \lambda(t,t_0) C(t_0) u(t,t_0),
\end{equation}
and yielded results in Fig. \ref{fig:gevp}. Note that the lowest level in the rest frame is above the inelastic threshold, which means the extracted energies will not contribute to a conventional $2$-particle Lüscher's analysis. To further investigate larger correlator matrices and moving frames in the future, we will extend the computation to have all correlators defined on every time slice, which will enable a finer tuning of $t_0$. Increasing statistics will help getting better signal in correlators with high momentum injections.

\section{Conclusions and Outlook}

By studying several distillation setups for various $N_{\mathrm{vec}}$, we observed that the smearing corresponding to $N_{\mathrm{vec}}=64$ consistently momentum-projects single-particle operators at different moving frames, and that going above that value is not particularly beneficial at correlator level. Moreover, after defining a cost-normalized standard deviation, we noticed that the stochastic setups with an increasing number of diluted Laplacian sources and $N_{\eta}=2$ do not present better cost-signal than on the exact setup at $N_{\mathrm{vec}}=64$.

Using exact distillation at $N_{\mathrm{vec}}=64$, we have computed multi-hadron correlators (on every $2$nd time slice) suitable for a minimal $K \pi$ scattering study at physical pion mass. We have performed a simple GEVP from a $5 \times 5$ rest-frame correlator matrix which yielded sensible results at this stage. Sat as a next step, we will increase our statistics and perform the variational analysis (and subsequently the finite-volume analysis) for $n \times n, \ n \lesssim 10$ correlator matrices with Dirac inversions done on every time slice at various moving frames. 

\vspace{1cm}

\textbf{Acknowledgements} 

The authors thank the members of the RBC and UKQCD Collaborations for the helpful discussions and suggestions.

N.L., F.E. and A.P. also kindly thank Mike Peardon for the invaluable discussions.

This work used the DiRAC Extreme Scaling service at the University of Edinburgh, operated by the Edinburgh Parallel Computing Centre on behalf of the STFC DiRAC HPC Facility (\href{https://dirac.ac.uk/}{\texttt{https://dirac.ac.uk/}}). The equipment was funded by BEIS capital funding via STFC grants ST/R00238X/1 and STFC DiRAC Operations grant ST/R001006/1. DiRAC is part of the National e-Infrastructure.

PB has been supported in part by the U.S. Department of Energy, Office of Science, Office of Nuclear Physics under the Contract No. DE-SC-0012704 (BNL). P.B. has also received support from the Royal Society Wolfson Research Merit award WM/60035.

N.L. \& A.P. received funding from the European Research Council (ERC) under the European Union’s Horizon 2020 research and innovation programme under grant agreement No 813942.

M.M. gratefully acknowledges support from the STFC in the form of a fully funded PhD studentship.

A.P. \& F.E. are supported in part by UK STFC grant ST/P000630/1. A.P. \& F.E. also received funding from the European Research Council (ERC) under the European Union’s Horizon 2020 research and innovation programme under grant agreements No 757646.

\bibliographystyle{JHEP}
\bibliography{/Users/s2000761/Documents/bib/library.bib}

\end{document}